\documentclass[]{aa}   

\usepackage{graphicx}
\usepackage{txfonts}
\usepackage{xcolor}
\newcommand{\refr}[1]{\color{black} #1}

\begin{document}

   \title{SimAb: A simple, fast and flexible model to assess the effects of planet formation on the atmospheric composition of gas giants} 
   


   \author{N. Khorshid
          \inst{1,2}
          \and
          M. Min
          \inst{1,2}
          \and
          J.M. D\'esert
          \inst{1}
          \and
          P. Woitke
          \inst{3,4,5}
          \and
          C. Dominik
          \inst{1}
          }

    \institute{Anton Pannekoek Institute for Astronomy, University of Amsterdam,
              Science Park 904, NL-1098XH Amsterdam, the Netherlands\\
              \email{n.khorshid@uva.nl}\\
         \and
             SRON Leiden,
             Niels Bohrweg 4,
             2333 CA Leiden, the Netherlands\\
        \and 
            Space Research Institute, Austrian Academy of Sciences,
            Schmiedlstrasse 6,
            A-8042 Graz, Austria\\
        \and
            SUPA, School of Physics \& Astronomy, St Andrews University,
            North Haugh,
            St Andrews, KY169SS, UK\\
        \and
            Centre for Exoplanet Science, University of St Andrews,
            North Haugh,
            St Andrews, KY169SS, UK
             }

   \date{Received Month 00, 2021; accepted Month 00, 2021}

 
  \abstract
   {The composition of exoplanet atmospheres provides us with vital insight into their formation scenario. Inversely, planet formation processes shape the composition of atmospheres and imprint their specific signatures. In this context, models of planet formation containing key formation processes help supply clues on how planets form. This includes constraints on their metallicity and Carbon-to-Oxygen ratio (C/O). Gas giants in particular are of great interest due to the amount of information we can achieve about their atmospheric composition from their spectra, and also due to their relative ease in observation.}
   { We present a basic, fast, and flexible planet formation model, called SimAb (Simulating Abundances), to form giant planets and study their primary atmospheric composition soon after their formation. }
   {In SimAb we introduce parameters to simplify the assumptions about the complex physics involved in the formation of a planet. This approach allows us to trace and understand the influence of complex physical processes on the formed planets. In this study we focus on four different parameters and how they influence the composition of the planetary atmospheres. These parameters are the initial protoplanet mass, the initial orbital distance of the protoplanet, the planetesimal fraction in the disk, and the dust grain fraction in the disk.}
   {We focus on the C/O ratio and the metallicity of the planetary atmosphere as an indicator of their compositions. We show that the initial protoplanet core mass does not influence the final composition of the planetary atmosphere in the context of our model. The initial orbital distance affects the C/O ratio due to the different C/O ratios in the gas phase and the solid phase at different orbital distances. Additionally, the initial orbital distance together with the amount of accreted planetesimals cause the planet to have sub-solar or super-solar metallicity. Furthermore, the C/O ratio is affected by the dust grain fraction and the planetesimal fraction. Planets that accrete most of their heavy elements through dust grains will have a C/O ratio close to the solar C/O ratio, while planets that accrete most of their heavy elements from the planetesimals in the disk will end up with a C/O ratio closer to the C/O ratio in the solid phase of the disk.}
   {By using the C/O ratio and metallicity together we can put a lower and upper boundary on the initial orbital distance where super-solar metallicity planets are formed. We show that planetesimals are the main source for reaching super-solar metallicity planets. On the other hand, planets that mainly accrete dust grains will show a more solar composition. Super-solar metallicity planets that initiate their formation farther than the CO ice line have a C/O ratio closer to the solar value.}

    \titlerunning{SimAb: Simulating abundances of the atmosphere of gas giant based on their formation}
   \maketitle
%

\section{Introduction}
\label{sec: introduction}
    Over the past few decades different groups addressed the question of whether the atmospheres of planets can provide any information on their formation histories. \citet{Mordasini_2016} showed that the formation history of a planet affects the composition of the planetary atmosphere.
    Further studies show that the composition of the atmosphere of planets is affected by the composition of the disk \citep{Oberg_2011, Helling_2014, Notsu_2020}. However, the composition of the disk can evolve over time. This evolution can affect the composition of the planets that are forming in these disks \citep{AliDib_2014, Piso_2016, Ilee_2017, Booth_2019}. On the other hand the processes that deliver solids to the planet forming envelope can play a considerable role in the composition. These processes can be either due to pebbles or planetesimals \citep{Booth_2017, Venturini_2016, Madhusudhan_2017, Voelkel_2020}. Additionally, \citet{Madhusudhan_2014}, showed that the composition of planetary atmospheres can provide us with information regarding the migration of the planet.

    In recent years, advancements in observation techniques \citep{Snellen_2010, Otten_2020, Brogi_2019} have led to characterizations of exoplanet atmospheres, primarily Hot Jupiters such as \citet{Hoeijmakers_2018, Helling_2014, Kesseli_2021, Serindag_2021}. 
    In some of these studies the connection between the formation history of a planet and the observation of the planet atmosphere is addressed \citep{Molliere_2020, Giacobbe_2021}. In these studies, they looked at the C/O ratio in the atmosphere of HD 209456b and HR 8799e. Based on the C/O ratio of HD 209456b they suggest the planet is most likely formed farther than the H$_2$O ice line and probably between the CO$_2$ and CO ice lines. The C/O ratio and the individual abundances of the carbon and oxygen in HR 8799e, suggest that the planet has accreted a solid mass between 65 to 360 earth masses and is more likely for the planet to have initiated its formation beyond the CO ice line. Advancements in instrument precision and wave band coverage in the future with JWST and ARIEL \citep{ARIEL} 
    makes it possible to test and study the theories that have been developed and validate the connections between the atmosphere composition to their formation history.

    Of all the elements in exoplanet atmospheres, carbon, oxygen, and their ratios are some of the most widely studied, where important connections to a planet's formation history have been drawn in various studies \citep{Oberg_2016, Cridland_2016, Notsu_2020}. Further studies have delved into other elements, such as nitrogen \citep{Bosman_2019, Cridland_2020_n2, Turrini_2020}. \refr{These} studies show by including the information we get about the nitrogen abundance in the giant planets we are able to get more information about planets that are formed beyond the CO snow line. Furthermore we can get information regarding the chemical evolution of the disk. Other studies have looked into the relative metallicity between a planet atmosphere and the host star and how the metallicity of the host star affects the metallicity of the planets \citep{Thorngren_2016, Santos_2017, Hasegawa_2019, Thiabaud_2015}. These studies showed that there is a direct relation between the composition of the star and the planets, however, there is not a direct relationship between the C/O ratio of the planets and the metallicity of the hosting star.

    The C/O ratio is a powerful tool to study the composition of the planetary atmospheres because of its connection to the planet formation history and its relative ease in observation. Nonetheless, the C/O ratio alone cannot help us distinguish between the many processes that affect it, including but not limited to migration, the evolution of the disk, and the late enrichment of the atmosphere by planetesimals. Other markers must be used to detangle the history of the planet in combination with the C/O ratio.

    In this study we construct a new way to look at planet formation that combines many vital processes together to form a gas giant planet. To do so, we bring the physical processes back to their essential form and dependencies. This allows us to consider all dependencies and parameterize unknown factors in the formation process. With SimAb\footnote{https://github.com/nkhorshid/SimAb} we can perform large scale parameter studies and eventually implement this into retrieval and atmospheric analysis tools.

    In section \ref{sec: method}, we discuss the basic assumptions. We describe how we set up the abundances in the disk in \ref{sec_1: chem}. In \ref{sec_1: disk model}, we explain how we set up the disk module. In section \ref{sec_1: atmosphere}, we describe how the forming planet accretes gas and solid from the disk. In section \ref{sec_1: migration}, we explain how SimAb takes migration into account. In section \ref{sec: pss}, we go over the how the parameters have been set up in this study. In section \ref{sec: results} we describe the results of SimAb and how the input values can affect them. In section \ref{sec: discuss} we discuss how our results in the context of previous studies can improve our understanding of how planet formation can be traced through the composition of the planetary atmospheres. Finally in section \ref{sec: conclusion} we provide a summary of this study.

 \section{Method}
\label{sec: method}
	We construct a planet formation simulation that takes planet core parameters, disk parameters, and the disk \refr{elemental abundances} as input, and returns the planet atmospheric \refr{elemental abundances} of a formed planet at a given orbital distance and a given mass. In this work, we simulate the formation of gas giants assuming core accretion and \refr{Type II migration}. By tracking the material the planet accretes, we track the atmospheric elemental abundances of planets. The simulation monitors the effects of the initial core mass, orbital separation, the dust grain fraction of the disk's solid phase, and the planetesimal fraction of the disk's solid phase.
	
\begin{figure*}[!htbp]
		
		\centering
		\includegraphics[height=7.5 cm]{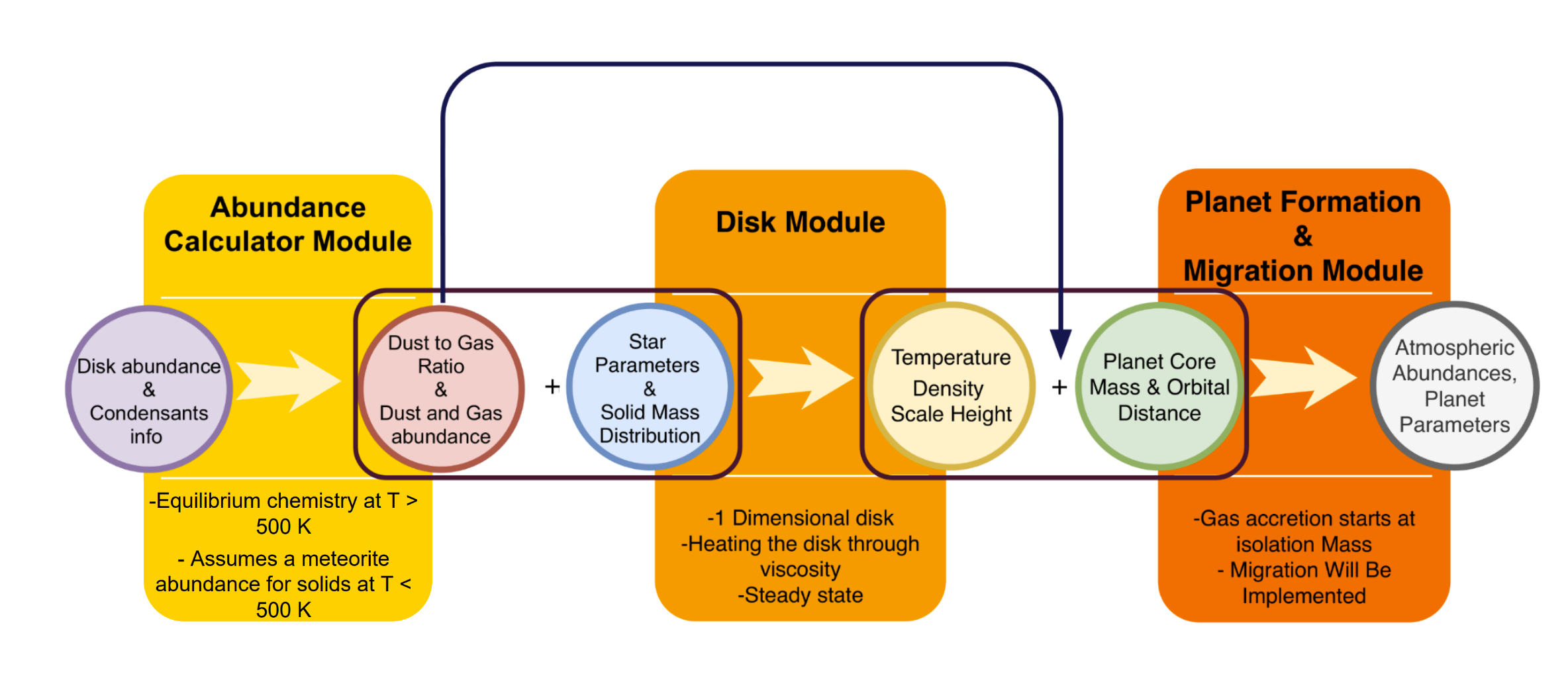}
		\caption{This figure shows a schematic of how SimAb functions. SimAb has three main modules: the abundance module, the disk module, and the planet formation and migration module. The abundance module takes the total abundance of the disk and the condensation temperature of the most abundant molecules that each atom can form at each temperature. Then based on the assumptions we integrated in it, calculates the dust-to-gas ratio and the abundances of the solid and gas for different temperatures with regard to the condensation temperatures. Additionally, the disk module calculates the disk temperature using a fixed dust-to-gas ratio and an alpha parameter, while it calculates the gas density using the dust-to-gas ratio that the abundance module calculates. The planet module takes the output of both the abundance module and the disk module together with the initial conditions of the protoplanetary core and calculates the mass and the composition of the planet based on a new dynamical alpha parameter to calculate the required migration rate to form the desired planet.} 
		\label{Fig: method}
	\end{figure*}

	SimAb is comprised of three main modules: A steady-state disk module, an abundance module, and a planet formation module. The overall structure of the method is depicted in figure~\ref{Fig: method}. The disk module indicates the properties of the protoplanetary disk such as its temperature, density, and scale height. We use an abundance module to compute the abundance of the atoms in the solid and gas phase. The abundance calculator then calculates the dust-to-gas ratio based on how much material is in the solid phase at each temperature. The planet formation module computes how much mass is accreted onto the planet from the solid and gas phases of the disk. Moreover, the planet formation module calculates the migration distance due to \refr{Type II migration}. In this simulation we adjust the migration speed by introducing a varying alpha parameter so the planet can reach the final orbital distance at the same time it reaches the desired mass. This allows us to form a planet with the same mass and orbital distance with various initial conditions. In the following sections we describe each of these modules and the assumptions therein.
	
	\subsection{Calculating the solid and gas abundances in the disk}
	\label{sec_1: chem}

    	The abundance module calculates the abundances of the different atoms in the solid and gas phases in the disk at a given temperature. Then, based on this composition, it calculates a dust-to-gas ratio. \refr{The values used in this study are not unique and there are many ways to deviate from the abundances assumed in this study. Therefore our model allows the use of other abundances for the atoms in the gas and solid phase. This abundance would heavily affect the results and should be taken into account when studying the results. In this study we use certain assumptions for the abundances of the atoms in the disk which we explain in this chapter.}
    	The disk is assumed to have a solar composition following the measurements by \citet{Asplund_2009}. Furthermore, we assume the condensation temperature is independent of the pressure and density in the disk. With these assumption, the module calculates the gas and solid abundance for each element for different temperatures in the disk. The atoms used in this work are presented in table~\ref{tab:Disk_compo}. We chose these atoms to keep the disk characterisation simple and have a good variety of atoms for further studies on planetary atmospheres. \refr{The composition of the disk is known to follow a disequilibrium chemistry. However, this is not true for a region close to the star within a few tenths of an AU. In this region the temperature and the density are higher, therefore  the composition can follow an equilibrium chemistry, following studies by \citet{Eistrup_2018, Oberg_2020}}. In the inner region with a temperature of over 500 K, corresponding to a distance of 0.7 AU, we use condensate predicted by equilibrium chemistry for the temperatures and typical densities of this region. For temperatures lower than 500 K, we assume that \refr{atoms that are used to form refractories are all in the solid phase. Therefore their abundances in the solid phase is the same as solar abundances. For other atoms, namely helium, carbon and nitrogen, we use meteorite abundances measured by \citet{Asplund_2009}. The meteorite abundance in this study is water rich therefore we cannot separate the abundance of oxygen that was added because of water ice or other forms. Therefore, we assume the Oxygen abundance at 500 K is the same as was calculated through the equilibrium chemistry. In order to determine how much hydrogen is in the solid phase we exclude the hydrogen abundance needed to form water from the hydrogen abundance that is reported in \citet{Asplund_2009}. At} each ice line the condensed species abundance is added to the solid state abundances. Using \refr{the assumption }in \citet{Oberg_2011}, we assume 15\% of the \refr{remaining} carbon locks up in CO$_2$ gas and the rest forms quickly into CO with the excess Oxygen forming H$_2$O.
	
        	\begin{table*}[ht!]
		        \begin{center}
			        \begin{tabular}{ |c|c|c| } 
				        \hline
				        Species         & $T_{cond}$\,[K] & limiting atomic abundance \\
				        \hline
				        \refr{Meteorite}      & 500   & \refr{as is described in section~\ref{sec_1: chem}} \\
				        NaAlSi$_3$O$_8$ & 850   & all remaining [Na], [Al], [Si], [O] \\
				        KAlSi$_3$O$_8$  & 850   & all remaining [K], [Al], [Si], [O] \\
				        MgAl$_2$O$_4$   & 1250  & $50\%\times$ remaining [Al] \\
				        Al$_2$O$_3$     & 1550  & all remaining [Al], [O]  \\
				        MgSiO$_3$       & 1150  & 83\%$\times$ remaining [Si]; all remaining [Mg] \\
				        Mg$_2$SiO$_4$   & 1200  & all remaining [Mg], [Si] \\
				        FeS             & 650   & all remaining [S] \\
				        Fe              & 1200  & all remaining [Fe] \\
				        H$_2$O          & 120   & all excess [O] \\
				        CO$_2$          & 47    & \refr{$15\%\times$remaining[C]$^{[1]}$}\\
				        CO              & 20    & all remaining [C] \\
				        He              & --    & \\
    				    \hline
				
	    		    \end{tabular}
		        	\caption{This chart shows the condensates we use in our abundance module and their condensation temperatures. These species are chosen since they are the most abundant species in an equilibrium chemistry at high temperatures based on equilibrium models. This chart also shows how much of the limiting atom is condensed in which form and at which temperature. This chart is organized as a task list. We work from top to bottom of this table. At each given temperature, we start from the first row that has a temperature higher than the given temperature and remove atoms from the gas phase. For each row we always check for the atom that limits the amount of condensate that can form. Then we condense all of that atom and all the other required atoms to form the condensate. If there are multiple condensates of the same atom for any given temperature, we use the percentage value given in their respective rows.[1]We use \citet{Oberg_2011} approach to determine the amount of carbon in CO$_2$ and CO. \refr{[2]We use \citet{Asplund_2009} to determine the meteorite composition.}}
			        \label{tab:Disk_compo}
		        \end{center}
		    
	        \end{table*}

\begin{figure}[!htbp]
		    \centering
		    \includegraphics[height=4.1cm]{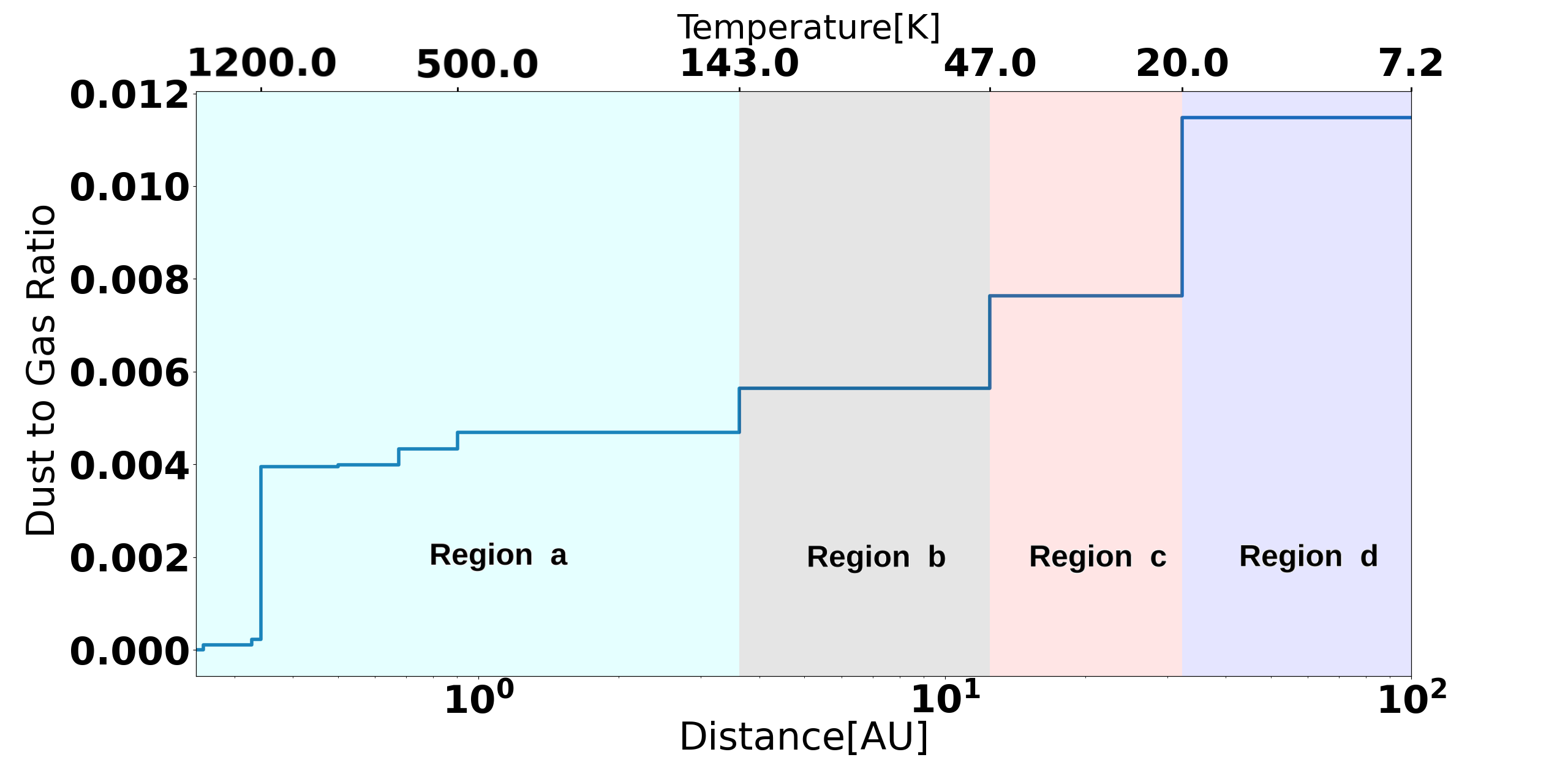}
		    \caption{The abundance module calculates the dust-to-gas ratio based on the amount of atoms in the solid and gas phase. This plot shows how the dust-to-gas ratio changes throughout the disk. The steps in the plot are caused by the condensation of different species mentioned in table (\ref{tab:Disk_compo}). The colors in the plot indicate different regions in the disk. The cyan color, region 'A', shows the disk within the water ice-line. The black color, region 'B', shows the disk farther than the water ice-line and within the CO$_2$ ice-line. The red color, region 'C', shows the disk between the CO$_2$ ice-line and CO ice-line. The blue color, region 'D', shows the disk farther than the CO ice-line. }
		
		    \label{Fig: dTg}
	    \end{figure}

\begin{figure}[!htbp]
		    \centering
		    \includegraphics[height=4.1cm]{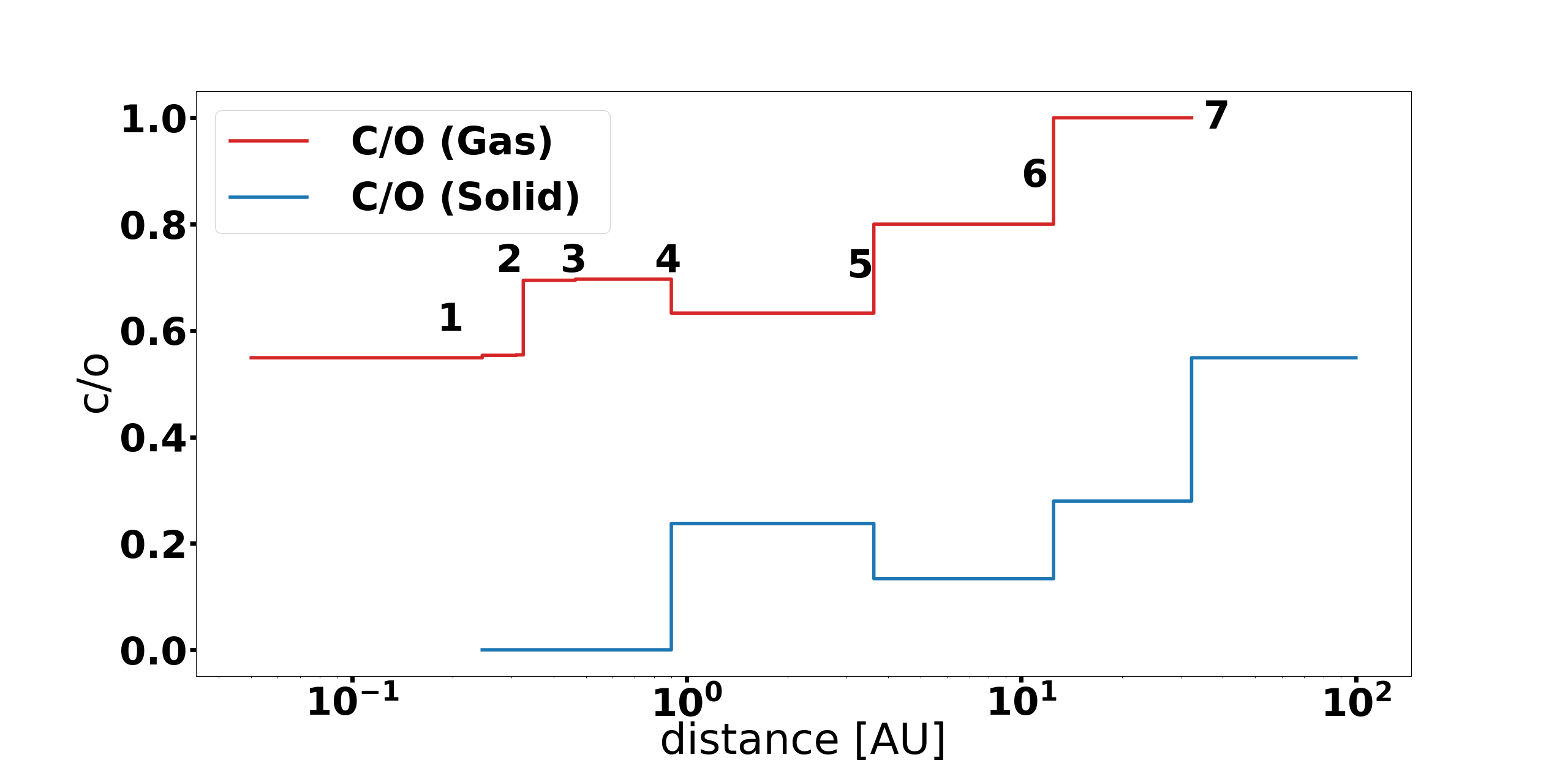}
		    \caption{The C/O ratio over the disk in the gas phase and solid phase. The C/O ratio starts as a solar value in the gas phase, closer to the star. By condensation of Al$_2$O$_3$, at point 1, Mg$_2$SiO$_4$ at point 2, and KAlSi$_3$O$_8$ and NaAlSi$_3$O$_8$ at point 3, the C/O ratio increases in the gas phase. Due to the lack of carbon in the solid phase for these steps, the C/O ratio is zero until point 4 where part of the carbon is condensed to form a \refr{meteorite}-like composition. This results in a decrease in the C/O ratio in the gas phase. At point 5 the condensation of H$_2$O increases the C/O ratio in the gas phase\refr{. At point 6 CO$_2$ condenses into ice form. This rises the C/O ratio in the gas phase up to 1}. Finally at point 7, by condensing CO, the C/O ratio becomes solar in the solid phase. The gas phase at this point is oxygen and carbon free.}
		    \label{Fig: CtO}
	    \end{figure}

    	In the hot region, we use a similar approach as mentioned in \citet{Min_2010}, assuming each atom forms a few main condensates as table~\ref{tab:Disk_compo} provides. At each given temperature, we first determine the main condensates in that temperature, and deplete all the atoms into those condensates with regards to the limiting atom. We use GG-chem \citep{Woitke_2018} to find out the main species formed for a wide range of pressures ($10^{1}$ to $10^{-6} bar$) for temperatures below 500K. The procedure of how the \refr{abundance module calculates} the abundances of the solid using these abundances is explained in table~\ref{tab:Disk_compo}. The abundance module then calculate the dust-to-gas ratio. Figure~\ref{Fig: dTg} shows the dust-to-gas ratio that the abundance module calculates. The increase in the dust-to-gas ratio is due to the condensation of the condensates mentioned in table~\ref{tab:Disk_compo}. Figure~\ref{Fig: CtO} shows the C/O ratio in the gas phase and the solid phase. \refr{The partition of carbon and oxygen is also presented in table ~\ref{tab: cto}}
	    
	    \begin{table*}[ht!]
		        \begin{center}
			        \begin{tabular}{ |c|c|c|c| } 
				        \hline
				        \refr{Species}         & \refr{$T_{cond}$\,[K]} & \refr{[O]/[H]$\times10^{-4}$} &\refr{[C]/[H]$\times10^{-4}$} \\
				        \hline
				        \refr{Refractories}      & \refr{>500}   & \refr{$1.04$} &\refr{$0$} \\
				        \refr{Meteorite}      & \refr{500}   & \refr{$0$} &\refr{$0.25$} \\
				        \refr{H$_2$O ice}      & \refr{120}   & \refr{$0.44$} &\refr{$0$} \\
				        \refr{C$_2$O ice}      & \refr{47}   & \refr{$1.95$} &\refr{$0.97$} \\
				        \refr{CO ice}      & \refr{20}   & \refr{$1.47$} &\refr{$1.47$} \\
    				    \hline
				
    	    		    \end{tabular}
		        	\caption{\refr{This table presents the partition of carbon and oxygen in the disk.}}
			        \label{tab: cto}
		        \end{center}
		\end{table*}
	
	\subsection{Disk Module}
	\label{sec_1: disk model}
	    We assume a one dimensional disk based on the analytical model represented in \citet{Min_2010}. The temperature is calculated assuming viscous heating, which is the main source of energy in the optically thick part of the disk. Equation~\ref{eq: temperatur} and equation~\ref{eq: density} calculate the temperature and density of the mid-plane:

	    \begin{equation}
	    \label{eq: temperatur}
	    T^{5} = \dfrac{3\mu m_p \Dot{M}}{3\pi^2\alpha k_b f \sigma}[\dfrac{GM_{\star}}{R^3}]^{3/2}
	    \end{equation}
	
	    \begin{equation}
	    \centering
	    \label{eq: density}
	    \Sigma_{gas} = \dfrac{\mu m_p \Dot{M}}{3\pi \alpha k_b T} \sqrt{\dfrac{GM_{\star}}{R^3}}
	    \end{equation}

	    In these equations, $\mu$ is the mean molecular mass of the gas, $m_p$ is the proton mass, $\Dot{M}$ is the mass accretion rate of the star, $\alpha$ is the viscosity constant, \refr{and $\frac{1}{f}$} is the dust-to-gas ratio. The dust-to-gas ratio varies between 0.01 to 0.004 depending on the atomic abundance of the solid and gas in the disk. We use a constant dust-to-gas ratio of 0.01 to calculate the temperature and a varying dust-to-gas ratio for the solid condensation. We assume that the planet is formed in a short time compared to the disk evolution so the star mass accretion rate as well as the ice line positions are constant during the planet formation, following \citet{Eistrup_2018}.

	    \begin{figure}[!htbp]
		
		    \centering
		    \includegraphics[height=4.2cm]{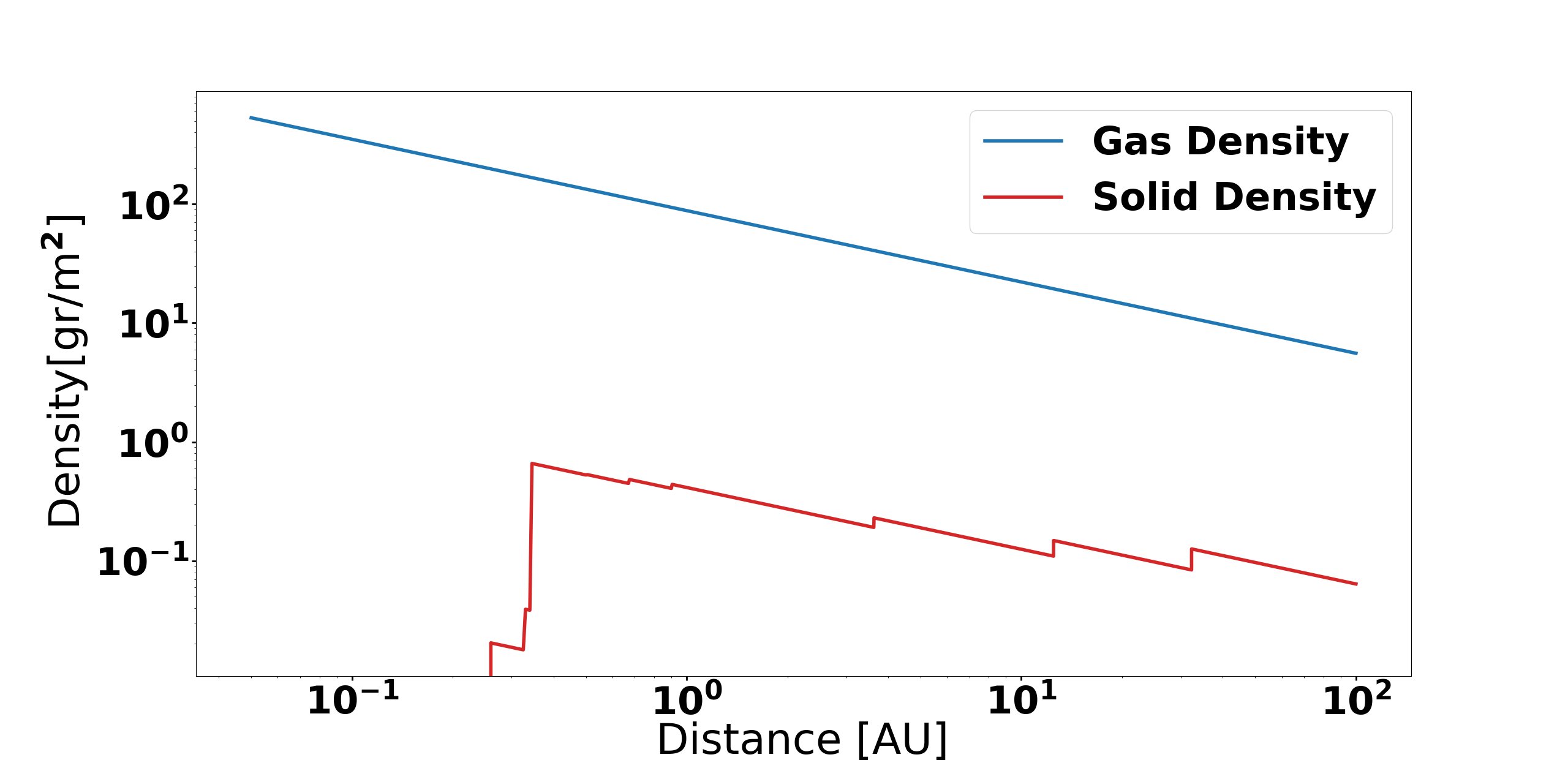}
		    \caption{This plot shows the density change of the gas and the solid at different orbital distances. The steps in the solid density is caused due to different dust-to-gas ratios at different orbital distances.}
		    \label{Fig: disk}
	    \end{figure}
	
	    In addition to the temperature and density, the disk module calculates the scale height of the disk $H$, the speed of sound $c_s$, and the angular velocity at the given orbital distance.

    \subsection{Planet Module}
	\label{sec_1: atmosphere}
	
	The planet module forms a gas giant planet through a core accretion model. In SimAb we assume each single simulation forms a single planet. Since we are interested in the atmospheric composition of the planets, SimAb starts the formation after the protoplanet core is formed and starts the runaway gas accretion. Furthermore we assume the protoplanet mass is larger than the pebble isolation mass. For simplicity, planets formed go through \refr{Type II migration} throughout the time they accrete their atmospheres. The model is set with the assumption that migration and gas accretion halt simultaneously, when the planet reaches the desired mass and orbital distance.

    The planet module calculates the composition of the atmosphere of planets, the mass of material accreted through the solid phase and the gas phase, as well as adjusting the migration speed of the planet so that the planet reaches the required mass and orbital distance at once. 

	    \subsubsection{Gas Accretion}

	    \label{sec_2: gas}
	        In SimAb the planet formation starts by assuming a core of a few Earth masses starting runaway gas accretion. This mass is already larger than the pebble isolation mass.
	
	        We use \citet{Bitsch_2019a} to calculate the accreted gas onto the planet. Equation~\ref{eq: gas_acc} shows the rate at which the planet accretes mass. The planet gas accretion rate is restricted by the Bondi accretion rate\citep{Ida_2004} and the viscous flux of the disk. In equation~\ref{eq: gas_acc}, the first part calculates the viscous flux and the result is equal to the star mass accretion, while the second part is the Bondi rate.
	        
	        \begin{equation}
	        \label{eq: gas_acc}
	        \dot{M}_{gas} = H^2 \Omega \Sigma_{g} \min\left(3\pi \alpha, (\dfrac{r_h}{H})^{9/2}\right)
	        \end{equation}

	        The accreted gas consists of volatile material as well as some refractories in the form of dust grains. These grains are small enough that they are fully coupled to the gas and move around with the gas. The composition of the gas then depends on the dust-to-gas ratio as well as how much of the solid is in the form of dust grains, the dust grain fraction. In the case where all the material is in the form of dust, the dust grain fraction is one, the composition of the accreted gas would be solar. The abundance module calculates the dust-to-gas ratio that we use in this step for each given temperature. The densities of the gas and solids that we use in this step are given in figure~\ref{Fig: disk}. The gas accretion stops once the planet reaches its final orbital distance which happens at the same time that the planet reaches its final mass.

	    \subsubsection{Solid Accretion}

	    \label{sec_2: solid}

	        The solid component of the disk is defined by four parameters. The first parameter is the \refr{dust to gas ratio, $\frac{1}{f}$,} which indicates what fraction of the disk mass is in solids. We introduce the dust grain fraction as the second parameter, $\delta_{dust}$, the pebble fraction as the third, $\delta_{pebble}$, and the planetesimal fraction as the last, \refr{$\delta_{pls}$}. These last three parameters are smaller than one and together add up to one. A ratio of zero defines a situation where there exists either no dust, no pebbles, or no planetesimals in the system (depending on which solid fraction has a ratio of zero) and a value of one represents the opposite case where the system is comprised solely of dust, pebbles, or planetesimals (depending on which solid has a fraction of one). In order to keep the model simple, we assume the solid fractions are the same through out the disk.
	        
	        We assume the starting core mass in our simulation is higher than pebble isolation mass, therefore the planet does not accrete the pebbles. On the other hand the simulation recognizes accretion of dust grains and planetesimals.
	        
	        Dust grains have a Stokes number much lower than one, are coupled to the gas, and are accreted through the runaway gas accretion. The dust grain mass accreted via runaway gas accretion is determined by the \refr{dust-to-gas ratio, $\frac{1}{f}$,} as well as the dust grain fraction, $\delta_{dstg}$. Equation~\ref{eq: dust mass} calculates the dust mass present in the accreted gas mass, $M_{acc,g}$.
	
	        \begin{equation}
	        \label{eq: dust mass}
	        M_{dstg} = \frac{\dot{M}_{g}}{f}~\delta_{dstg}
	        \end{equation}

	        Planetesimals are defined as objects with a Stokes number much higher than one and have diameters around a few km to a few hundreds of km. Planetesimals are not much affected by the gas drag and even after the gap opening, planets can accrete planetesimals. We assume that the planetesimals are spread throughout the disk with the same distribution as the gas density. The dynamical interactions between the planet and the planetesimals is ignored. The planet accretes planetesimals present in its feeding zone, which we define as one Hill radius around the planet. However, for the purpose of our study the exact radius of the feeding zone does not matter as the planet will eventually accrete \refr{a fix ratio of} the planetesimals in the area that it swipes through its formation\refr{,~$\epsilon_{pls}$. This ratio together with the planetesimal fraction, in this model, is called the planetesimal ratio,~$\delta_{\epsilon,pls}$} Equation~\ref{eq: planetesimal mass} shows the increase of planet mass by accreting planetesimals:
	
	        \begin{equation}
	        \label{eq: planetesimal ratio}
	        \delta_{\epsilon,pls} = \delta_{pls} \times \epsilon_{pls}
	        \end{equation}
	        
	        \begin{equation}
	        \label{eq: planetesimal mass}
	        \dot{M}_{pls} = \frac{4\pi \times r \times \dot{r}_p \times \Sigma_{g}}{f}~\delta_{\epsilon,pls}
	        \end{equation}
	
	        In this equation, $\Sigma_{g}$ is the gas density, and $\delta_{\epsilon,pls}$ is the \refr{planetesimal ratio} and represents the planet efficiency in accreting planetesimals and the planetesimal mass ratio to the total solid mass. The planetesimals are dissolved in the atmosphere and the remaining mass reaching the surface of the planet core is negligible \citep{Podolak_1988, Pollack_1996, Venturini_2016}.
	
	        The planetesimal composition is dependent on their formation region therefore SimAb allows for different assumptions for the planetesimal composition. In this study we adjust the composition of the planetesimals to the composition of the solids at the distance they are accreted.

    \subsection{Migration and Evolution}
	\label{sec_1: migration}
	
	    SimAb assumes the planet is only going through \refr{Type II migration} with an inward direction through the disk. The migration stops when the planet reaches its final orbital distance, which is given as an input in the initial parameters. This assumption allows us to have an estimate of the composition of the planet without overly complicating the model.
	
    	We implement migration using the migration rate from \citet{Alibert_2005}, shown in equation~\ref{eq: mig_rate}. The migration rate is slower once the mass of the gas within the planetary orbit is less than the planet mass, however, we assume the same migration rate, even when the planet mass is heavier than the disk mass within its orbit. This assumption allows us to adjust the model in a way so that all the formed planets can end in the same spot, without affecting their accretion rate. Using an accretion rate that is dependent on the planet mass would interact with the computational aspect of the model, which forces the planet to finish its formation at the same orbital distance and mass, and would force the planet to accrete majority of its mass in-situ. We discuss the impact of such a simplification in Section \ref{sec: discuss}.
	
    	\begin{equation}
    	\label{eq: mig_rate}
    	\dot{r}_p = \dfrac{3\alpha_d H^2 \Omega}{2 r_p}
    	\end{equation}

    	Here, $\alpha_d$ is the dynamical viscous constant of the disk, $\Omega$ is the angular velocity of the disk, $r_p$ is the planet orbital distance, $M_p$ is the planet mass, and $\Sigma_g$ is the gas density. $\alpha_d$ can vary between $10^{-6}$ for very low viscosity disks, and less than $1$, for high viscosity disks. Varying $\alpha$ allows us to change the migration timescale and match it to the formation timescale of the planet. This assumption \refr{allows us to reproduce the formation of} hot and ultra hot giant planets from small cores as far as 100 AU or form planets in-situ.

    	SimAb is not robust against non-physical $\alpha_d$ values with values greater than one. For $\alpha_d$ values close to one some of the assumptions we make are no longer valid. Even though focusing on such planets is beyond the scope of this study, such values would allow for prediction of planets formed with a similar composition that must have migrated through methods aside from just disk viscosity migration.

	
\section{Setting up the variables}
\label{sec: pss}

	We run simulations for a hundred thousand different initial conditions 
	to form a Jupiter mass planet at 0.02 AU. There are four main parameters governing the formation of these planets: The initial orbital distance, the initial core mass, and the two separate parameters for solid size distributions: the dust grain fraction and the planetesimal fraction in the disk.

	The boundaries of these input parameters are given in table~\ref{tab:m_input}. The initial orbital distance cannot be closer to the star than the final orbital distance due to the model not supporting outward migration. The planet cannot start its runaway gas accretion at the same distance as the final orbital distance as the migration step size must be greater than zero. The initial mass can vary between five earth masses to thirty earth masses \citep{Piso_2014a, Piso_2014b}. We assume that the dust grain fraction and the planetesimal fraction is constant throughout the disk. The value of the other variables that we use in the model are given in table~\ref{tab:m_var}.

	\begin{table*}[ht!]
		\begin{center}
			\begin{tabular}{ |c|c|c|c| } 
				\hline
				Parameter         & Min & Max & Distribution \\
				\hline
				Initial Mass      & 5 $M_{\oplus}$  & 35 $M_{\oplus}$ & uniform distribution \\
				Initial orbital distance & 0.2 AU   & 100 AU & uniform distribution \\
				Planetesimal fraction  & 0   & 1 & Uniform distribution \\
				Dust grain fraction   & 0  & 1 & Uniform distribution \\
				\hline
				
			\end{tabular}
			\caption{This table shows the variable input parameters and their ranges.}
			\label{tab:m_input}
		\end{center}
		
	\end{table*}
	
	\begin{table*}[ht!]
		\begin{center}
			\begin{tabular}{ |c|c| } 
				\hline
				Parameter         & value \\
				\hline
				Mean molecular mass      & 2.5 \\
				alpha parameter & 0.1\\
				Rosseland mean opacity of the gas/ice mixture & 572 $cm^2/g$\\
				Final orbital distance  & 0.02 AU \\
				Final planet mass   & 318 $M_{\oplus}$  \\
				Stellar mass & 1 $M_{\odot}$\\
				Star mass accretion & $10^{-7} \dfrac{M_{sun}}{yr}$\\
				\hline
				
			\end{tabular}
			\caption{This table shows the fixed input parameters of the model.}
			\label{tab:m_var}
		\end{center}
		
	\end{table*}
	

\section{Results}
\label{sec: results}
\begin{figure*}[!htbp]
		\centering
		\includegraphics[height=10cm]{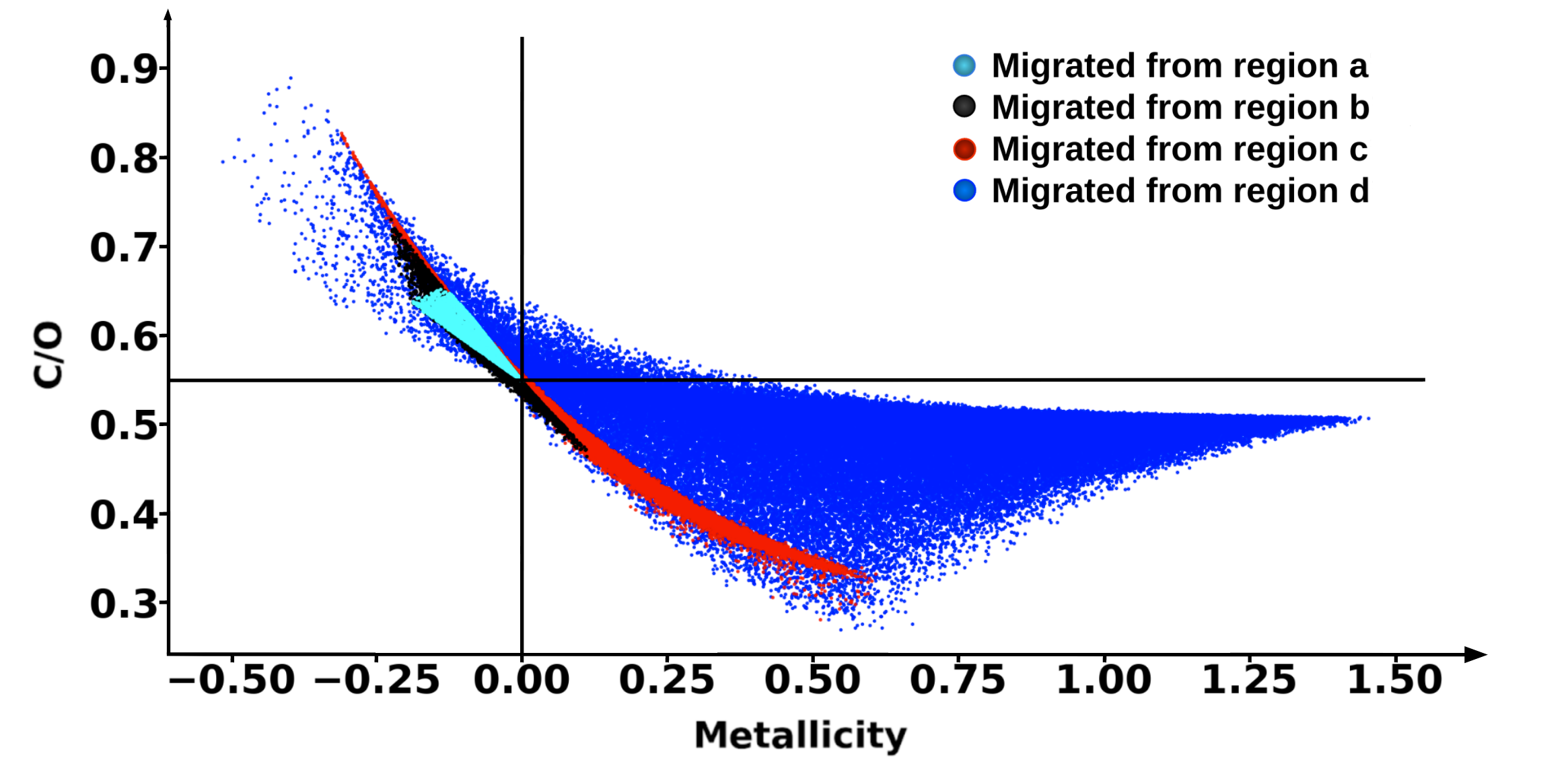}
		\caption{This figure shows the the C/O ratio and the \refr{log scale of} metallicity of \refr{a hundred} thousand planets with one Jupiter mass at a distance of 0.02 AU with different compositions. The different compositions are caused due to the different formation initial conditions that are explained in section \ref{sec: pss}. The five different colors shows planets that initiated their formation in five different regions of the disk. The cyan color, shows planets that initiated their migration within the water ice line, region 'A'. The black color, are planets that initiated their migration farther than the water ice line and within the CO$2$ ice line, region 'B'. The red color, are planets that initiated their migration beyond the CO$2$ ice line and within the CO ice line, region 'C'. The blue color, are planets that initiated their migration farther than the CO ice line, region 'D'.}
		\label{Fig: regions}
	\end{figure*}

    We ran simulations for a hundred thousand different initial conditions following the above procedure to form a Jupiter mass planet at a distance of 0.02 AU.	
	The calculated C/O ratio versus metallicity of the formed planets from this run are plotted in figure~\ref{Fig: regions}. The plot uses four different colors to encode the starting distance of the planet embryo. \refr{The metallicities we report in this work only includes the atoms we trace in SimAb and does not trace other atoms. This also present itself in solar metallicity. Moreover, the core composition does not contribute to the metallicity of the planets in this work.}

\begin{figure*}[!htbp]
		\centering
		\includegraphics[height=11cm]{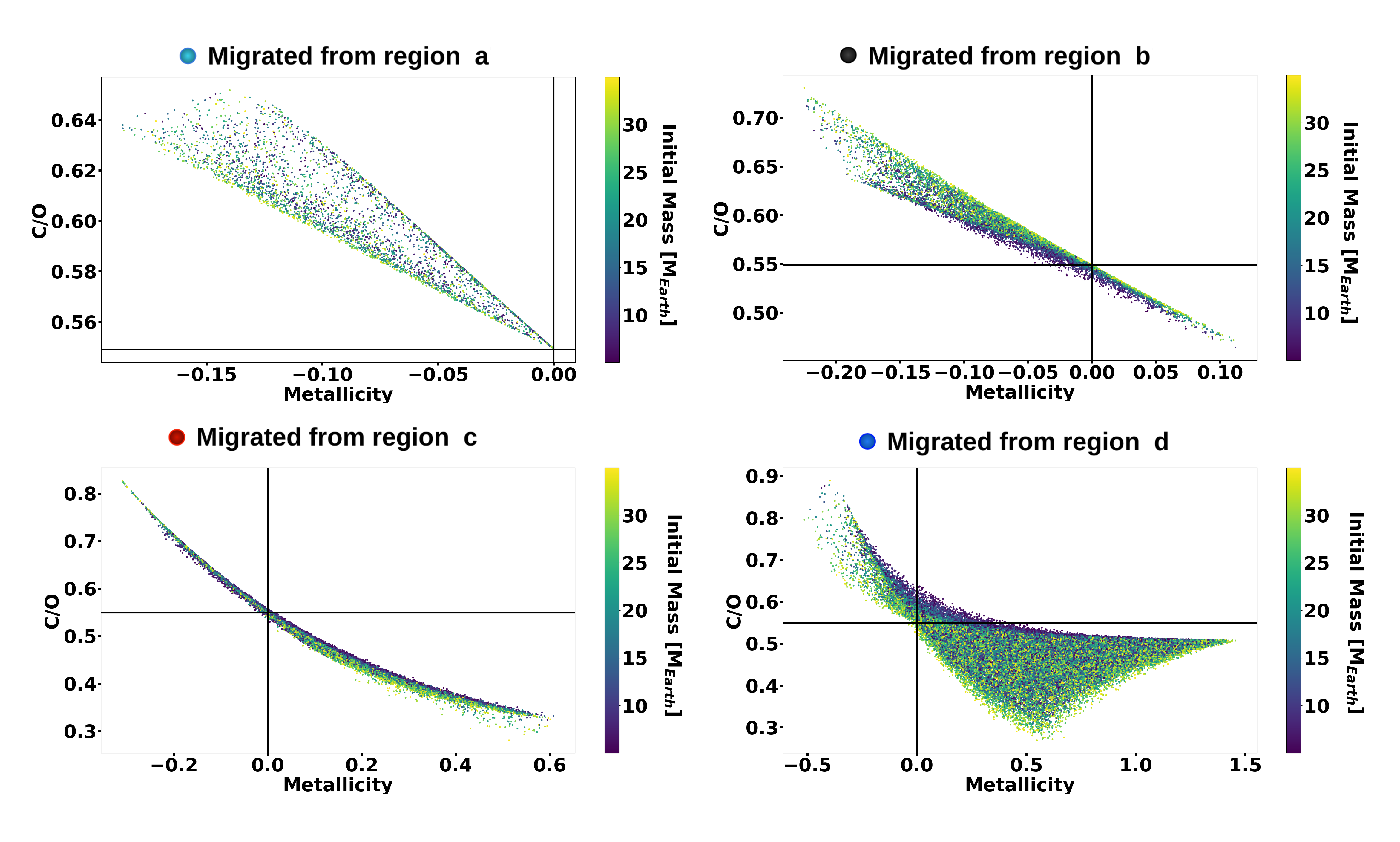}
		\caption{This figure shows the C/O ratio and \refr{the log scale} metallicity of the planets formed by the model and were previously presented in figure \ref{Fig: regions}. The color-bar indicates the initial core mass of these planets. We zoomed in on the four regions explained in figure~\ref{Fig: regions}. The top left panel shows planets that initiate their formation in region 'A', the top right panel shows planets that initiate their formation in region 'B', the bottom left panel shows planets that initiate their formation in region 'C' and the bottom right panel shows planets that initiated their migration from region 'D'. In all four panels there is no clear correlation between the C/O ratio and the metallicity of the planetary atmospheres and their initial core mass.}
		\label{Fig: mass}
	\end{figure*}

\begin{figure*}[!htbp]
		\centering
		\includegraphics[height=11cm]{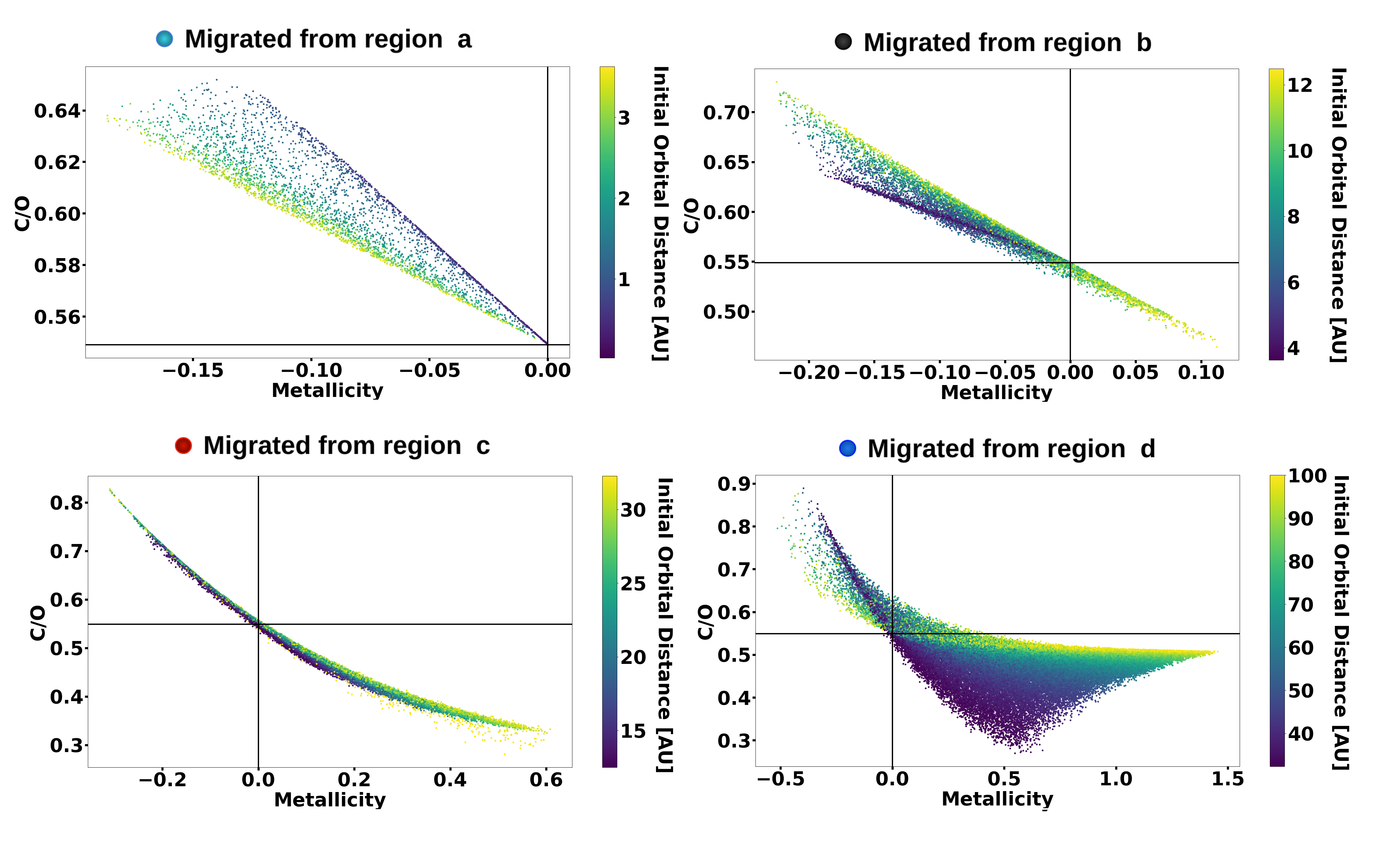}
		\caption{This figure shows the C/O ratio and \refr{the log scale }metallicity of the planets formed by SimAb and were previously presented in figure \ref{Fig: regions}. The color-bar indicates the initial core orbital distances of these planets. We zoomed in on the four regions explained in figure~\ref{Fig: regions}. The top left panel shows planets that initiate their formation in region 'A': in this plot we see that planets that initiate their formation farther out would end up with a lower C/O ratio comparably. The top right panel shows planets that initiate their formation in region 'B': In this region, planets with sub-solar metallicity that are formed farther out have higher C/O ratios. Additionally, only planets that are formed closer to the CO$2$ ice line show super-solar metallicity. The bottom left panel shows planets that initiate their formation in region 'C': The C/O ratio in these planets shows values closer to solar as they initiate their formation farther out. The bottom right panel shows planets that initiate their formation in region 'D': The C/O ratio shows values closer to the solar value as the planet initiates its formation farther out.}
		\label{Fig: distance}
	\end{figure*}

\begin{figure*}[!htbp] 
	
		\centering 
	
		\includegraphics[height=11cm]{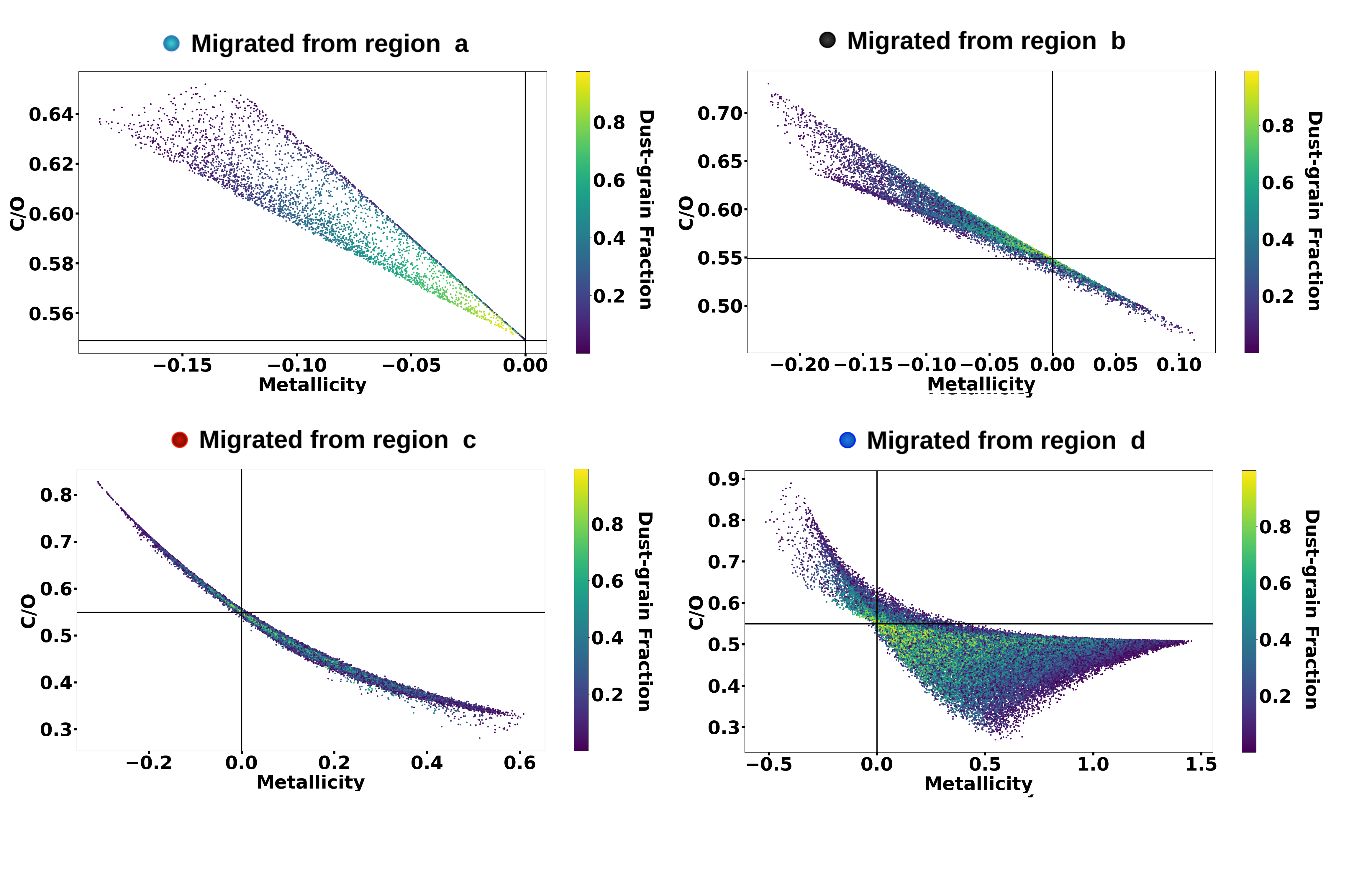} 
	
		\caption{This figure shows the C/O ratio and \refr{the log scale }metallicity of the planets formed by SimAb and were previously presented in figure \ref{Fig: regions}. The color-bar indicates the dust grain fraction. We zoomed in on the four regions explained in figure~\ref{Fig: regions}. In all four plots we see that planets that are formed in disks with higher dust grain fractions show metallicity and a C/O ratio closer to the solar value. This relation becomes less strong for planets that initiate their formation in farther regions, such as region 'C' and 'D'} 
	
		\label{Fig: dstg} 
	
	\end{figure*}

	Figures~\ref{Fig: mass}-\ref{Fig: dstg} show how each of these regions is affected by the four main parameters. In the coming paragraphs we explain these plots.
	
	Figure~\ref{Fig: mass} consists of four sub plots. These sub-plots show the C/O ratio versus metallicity of the atmosphere of the planets that SimAb produces. Each of the sub-plots is a zoom-in on a different region that was previously introduced in figure~\ref{Fig: regions}.
	
	The color-bar in these plots indicates the initial mass of the protoplanets. The initial core mass varies between 5 and 35 Earth masses. This plot shows that the composition of the planets does not have a clear correlation with the initial mass of the protoplanets. The lack of correlation between the initial mass and the composition of the planets is because metallicity and the C/O ratio are both affected by the amount of solids that the planet accretes. SimAb takes into account two ways to accrete solids onto the atmosphere: accretion of dust grains in the gas and accretion of planetesimals. The amount of dust grains in the gas is governed by the dust-to-gas ratio and how much solid mass is in the form of dust grains in the disk. On the other hand, in this model, the solid mass accreted through planetesimals does not depend on the initial mass of the planet.

	 Figure~\ref{Fig: distance} shows the relation between the composition of the planets, the C/O ratio and metallicity, and the initial orbital distance of the protoplanet. Each sub-plot shows planets that initiated their formation in a different region.
	 In this figure, the color-bar shows the initial orbital distance of the protoplanet core. All the sub-plots show a clear correlation between the composition of the planet atmosphere, mainly the C/O ratio, and where the planet initiates its formation. However, depending on which region the planet initiates its formation, these correlations have different characteristics. In order to understand why this relation changes for different regions, we look at the planets with sub-solar metallicity and super-solar metallicity separately for these regions. The reason for this approach is that planets with subsolar metallicity accrete most of their atmosphere mass from the gas phase of the disk while planets with super-solar metallicity accrete a substantial mass from the solid phase compared to the latter. Figure~\ref{Fig: CtO} shows the C/O ratio variations as a function of distance from the star, for both the solid phase and the gas phase in the disk. For planets with sub-solar metallicity we need to focus on the C/O ratio of the gas phase in the disk. For super-solar metallicity planets, we need to mainly focus on the C/O ratio of the solid phase in the disk.
	 
	 In the top left sub-plot, which is a zoom in on planets \refr{initiating} their migration from region 'A' in figure~\ref{Fig: regions}, the C/O ratio decreases as the planet's initial orbital distance increases. All of this region lies in the sub-solar metallicity half of the C/O-metallicity plot. By looking at the C/O ratio in the gas phase for distances within the water ice-line we see the C/O ratios increase for a short distance up to 0.7 AU. When the carbonaceous material condenses, this value decreases in the gas phase. Therefore planets that initiate their formation beyond 0.7 AU accrete part of their mass from a gas with a lower C/O ratio. The farther they initiate their formation, the more mass they accrete with a lower C/O ratio compared to planets initiating their formation in closer distances.

	 Planets that initiate their formation beyond the water ice line can have both super-solar and sub-solar metallicity. We first focus on the planets with sub-solar metallicity. In region 'B', the C/O ratio in the gas phase increases due to the condensation of the water into its icy form. Therefore planets that mainly accrete from the gas phase will end up with an increasing C/O ratio as their initiating distance increases. On the other hand, those planets that have a super-solar \refr{metallicity} tend to have a decreasing C/O ratio as well as increasing metallicity when increasing their starting distance from the star.

	 \refr{In region 'C', beyond the CO$_2$ ice line and within the CO ice line, the C/O ratio in both the solid and gas phase increases. In this region we see that planets that initiate their formation further out acquire a higher C/O ratio which is as expected. regardless of the phase of the material a planet accretes in this region, the further a planet  initiates its formation the more material it accretes with a higher C/O ratio. Although, planets with super-solar metallicity will have a higher C/O ratio compared to planets that initiate their formation closer to the star tend to have a much lower C/O ratio as their metallicity increases. These planets tend to accrete more from the solid phase and their C/O ratio approaches the C/O ratio in the solid phase.}
	 
	 
	 Beyond the CO ice line, all the carbon and oxygen will be in the solid phase. In this region, the solid phase will have a solar C/O ratio, while the gas phase is fully depleted from oxygen and carbon. Therefore the C/O ratio of the planets that mainly accrete gas is largely influenced by the C/O ratio in the gas phase within the CO ice line. Figure \ref{Fig: CtO} shows that the C/O ratio in the gas phase within the CO ice line is always super-solar, consequently the C/O ratio of such planets will be super-solar. On the other hand, planets that accrete a high mass from the solid phase (when the accreted solid mass to the accreted gas mass is higher than the dust-to-gas ratio of the disk) will end up with a sub-solar to solar C/O ratio. This is because the C/O ratio in the solid phase is sub-solar everywhere in the disk within the CO ice line, and is solar beyond the CO ice line. The farther out the planet initiates its formation, the more solid it can accrete with a solar C/O ratio. As a result, planets that initiate formation farther away from the star, will have a C/O ratio closer to the solar value.

	Figure~\ref{Fig: dstg} shows the relation between the C/O ratio and the metallicity of the atmosphere of the planets to the amount of solid mass that the planet accretes in the form of dust grains. The plot consists of four sub-plots which are oriented similar to the previous plots. The color-bar in these sub-plots shows how much dust is available for the planet to accrete. The yellow color indicates disks where the majority of the solid is in dust grain form, and dark blue indicates a disk where there is not much dust available in the disk. 

	The top left sub-plot, showing protoplanets that migrated from region A, shows a very strong correlation between the C/O ratio and metallicity and the amount of dust grains accreted onto the planet. The planets that are formed in disks with a higher abundance of dust are more likely to have a solar composition. On the other hand, planets that are formed beyond the CO$_2$ ice line do not show as much correlation. Nevertheless, planets that are formed in a disk with a high dust grain fraction tend to have a C/O ratio and a metallicity closer to solar compared to planets that are formed in disks where not much dust is available. As planets in this model accrete gas and dust simultaneously, in the extreme case where the dust grain fraction is one, the accreted material and thus the atmosphere of such planets will have a solar composition. This explains why increasing the amount of dust in the natal disk causes the formed planet to have a more solar composition.

\begin{figure*}[!htbp] 
	
		\centering 
	
		\includegraphics[height=11cm]{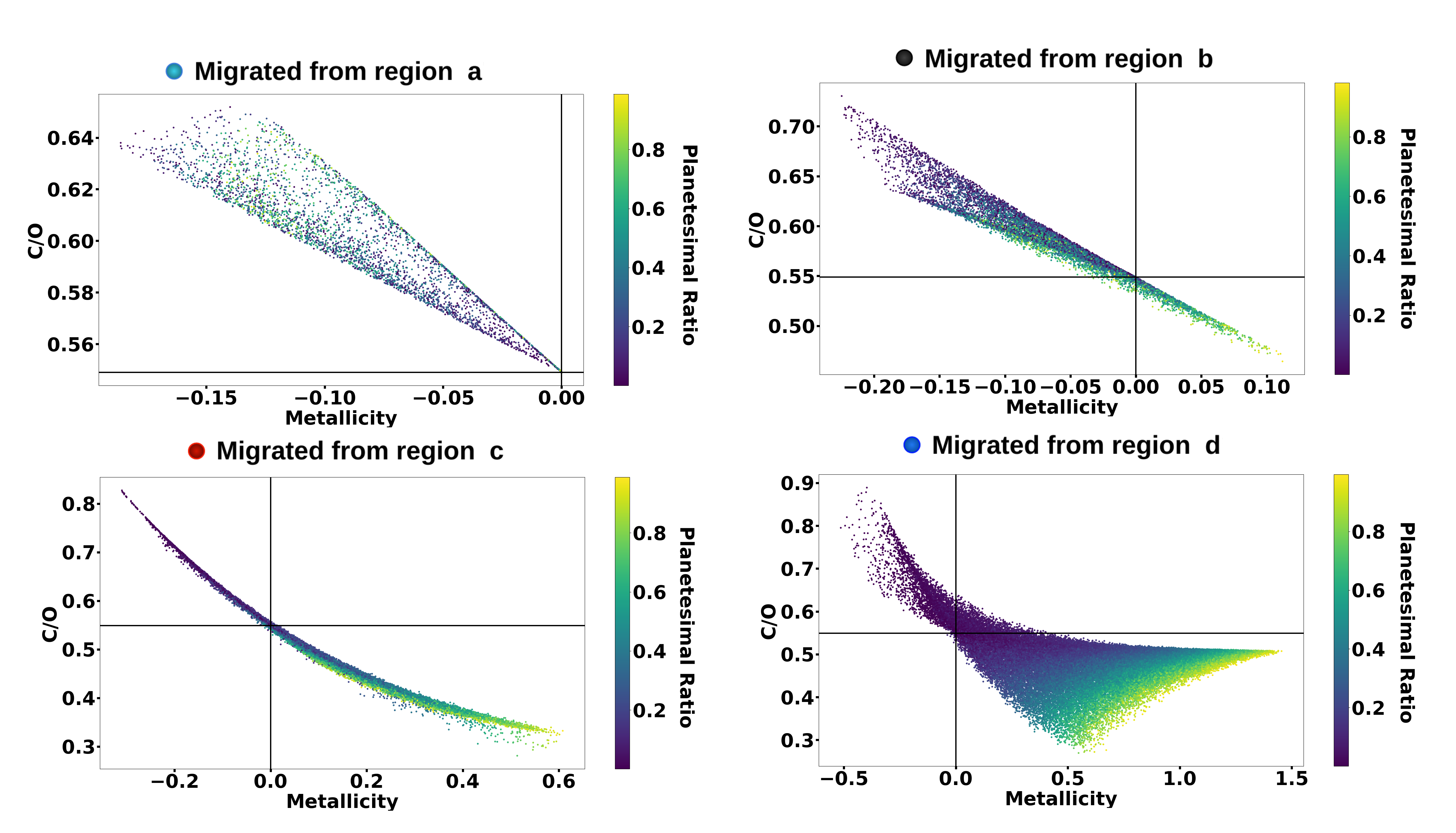} 
	
		\caption{This figure shows the C/O ratio and \refr{the log scale }metallicity of the planets formed by SimAb and were previously presented in figure \ref{Fig: regions}. The color-bar indicates the planetesimal fraction. We zoomed in on the four regions explained in figure~\ref{Fig: regions}. Planets that initiate their formation in region 'A' in disks with higher planetesimal ratio, show lower metallicity and a higher C/O ratio. On the other hand, planets that initiate their formation in regions 'B', 'C', or 'D' show a very strong correlation with the planetesimal fraction. Planets that initiate their formation in these regions with a high planetesimal fraction show a higher metallicity and a lower C/O ratio. In all four plots we see that the planets that are formed in disks with higher dust grain fractions show metallicities and C/O ratios closer to the solar value. This relation becomes less strong for planets that initiate their formation in farther regions, such as regions 'C' and 'D'} 
	
		\label{Fig: plnts} 
	
	\end{figure*}

	In figure~\ref{Fig: plnts}, similar to the previous plots, we plotted the C/O ratio versus the metallicity of the formed planets. The color-bar in these plots shows the overall efficiency of accreting planetesimals and how many planetesimals are available for the formation of each planet. \refr{Planets that are shown with the color yellow, are assumed to be very efficient in accreting planetesimals and are formed in disks with a high fraction of solid mass being in planetesimal form. On the other hand, planets that are shown with darker colors are formed in disks with not much planetesimals and or are not very efficient in accreting planetesimals.} This plot shows that planets that are formed within the water ice line, region 'A', behave differently compared to those that initiate their formation beyond the CO$_2$ ice line, regions 'C' and 'D'. The composition of these planets is mainly affected by the dust grain fraction, and the accreted planetesimals would not affect the composition. The correlation that we see in this plot is mainly coming from the assumption that the planetesimal fraction together with the dust grain fraction cannot acquire a value higher than one. Planets that initiate their formation within the water ice line show a weaker correlation between their composition, the C/O ratio and metallicity, and the planetesimal fraction. On the other hand, planets that are formed farther than the CO$_2$ ice line show a very strong correlation with the planetesimal fraction. Planets that are formed beyond the CO$_2$ ice line show a higher metallicity as their efficiency in accreting planetesimal increases. They generally show a lower C/O ratio compared to planets that are formed in the same distance. To understand the behavior of the composition of the planets regarding their efficiency in accreting planetesimals for those planets that are formed in region 'B', we need to understand at what distance the accretion of planetesimals can dominate the refractory composition of the planetary atmosphere. 

\begin{figure}[!htbp] 
		
		\centering 
		
		\includegraphics[height=5cm]{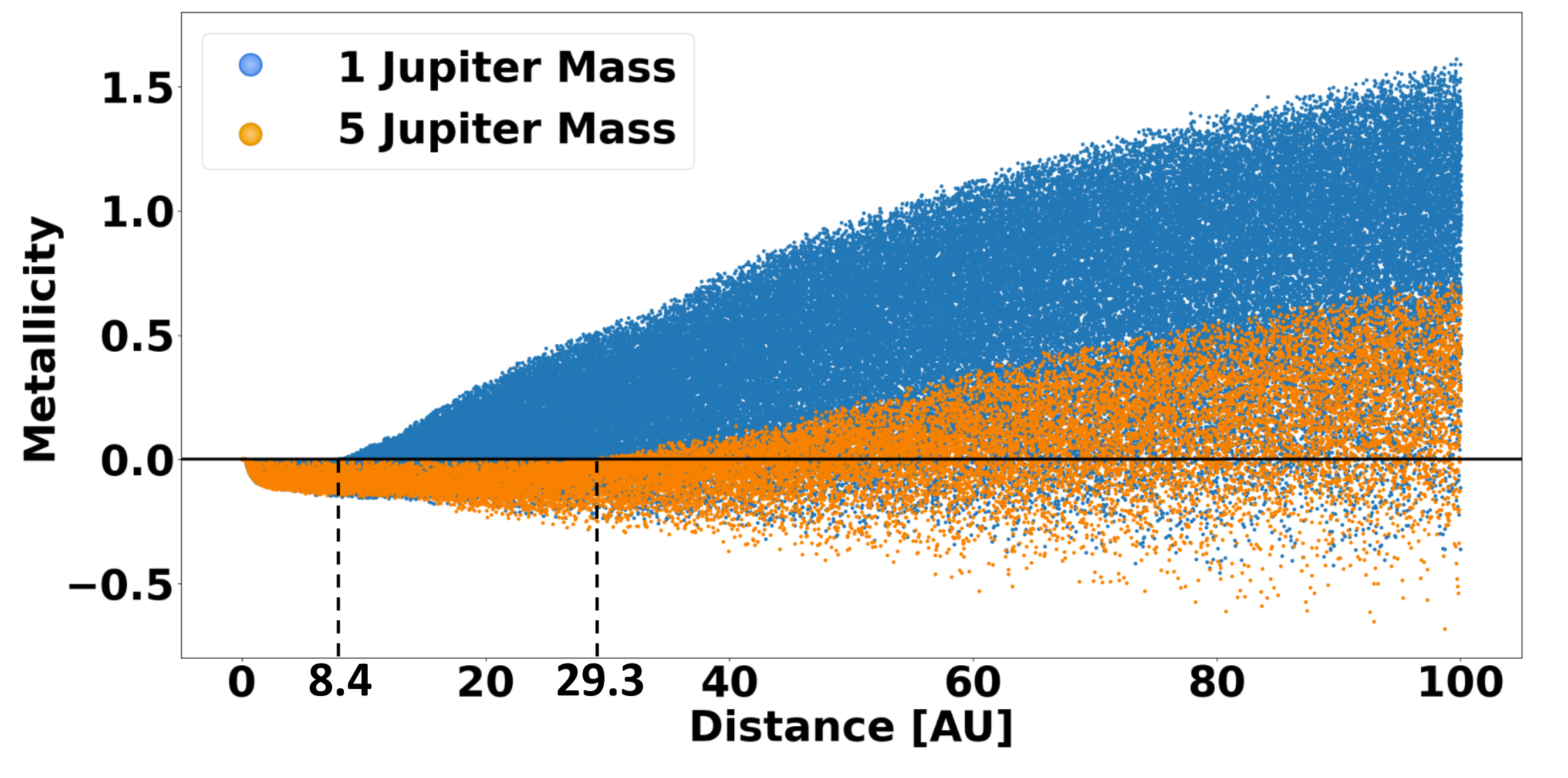} 
		
		\caption{This figure shows the \refr{log scale }metallicity and the initial orbital distance of planets of one Jupiter mass (blue) and five Jupiter masses (orange) at an orbital distance of 0.02 AU. This plot shows that planets with five Jupiter mass need to initiate their formation farther out compared to planets with one Jupiter mass in order to have a super-solar metallicity. The dashed lines shows the initial orbital distances that these planets start showing a super-solar metallicity.} 
		
		\label{Fig: met_dist} 
		
	\end{figure}

	Figure~\ref{Fig: met_dist} shows the metallicity of planets that SimAb forms compared to the distance where these planets initiate their formation. Planets that are shown in blue are Jupiter mass planets at a distance of 0.02 AU, while planets in orange have five Jupiter masses and are at the same orbital distance. \refr{This plot shows that heavier planets acquire lower metallicities compared to less massive planets when initiating their formation from the same distance. Hence }in order to form a heavy planet with super-solar metallicity, it must form farther out compared to a less massive planet. In addition to this, planets that are more efficient in accreting dust grains tend to have solar metallicity, as demonstrated by figure~\ref{Fig: dstg}, while planets that accrete more planetesimals end up with higher metallicity, as shown in figure~\ref{Fig: plnts}. From all this we can conclude that heavier planets need to accrete a higher mass in the form of planetesimals compared to less massive planets in order for their metallicity to be super-solar.

\begin{figure}[!htbp] 
	
		\centering 
	
		\includegraphics[height=5cm]{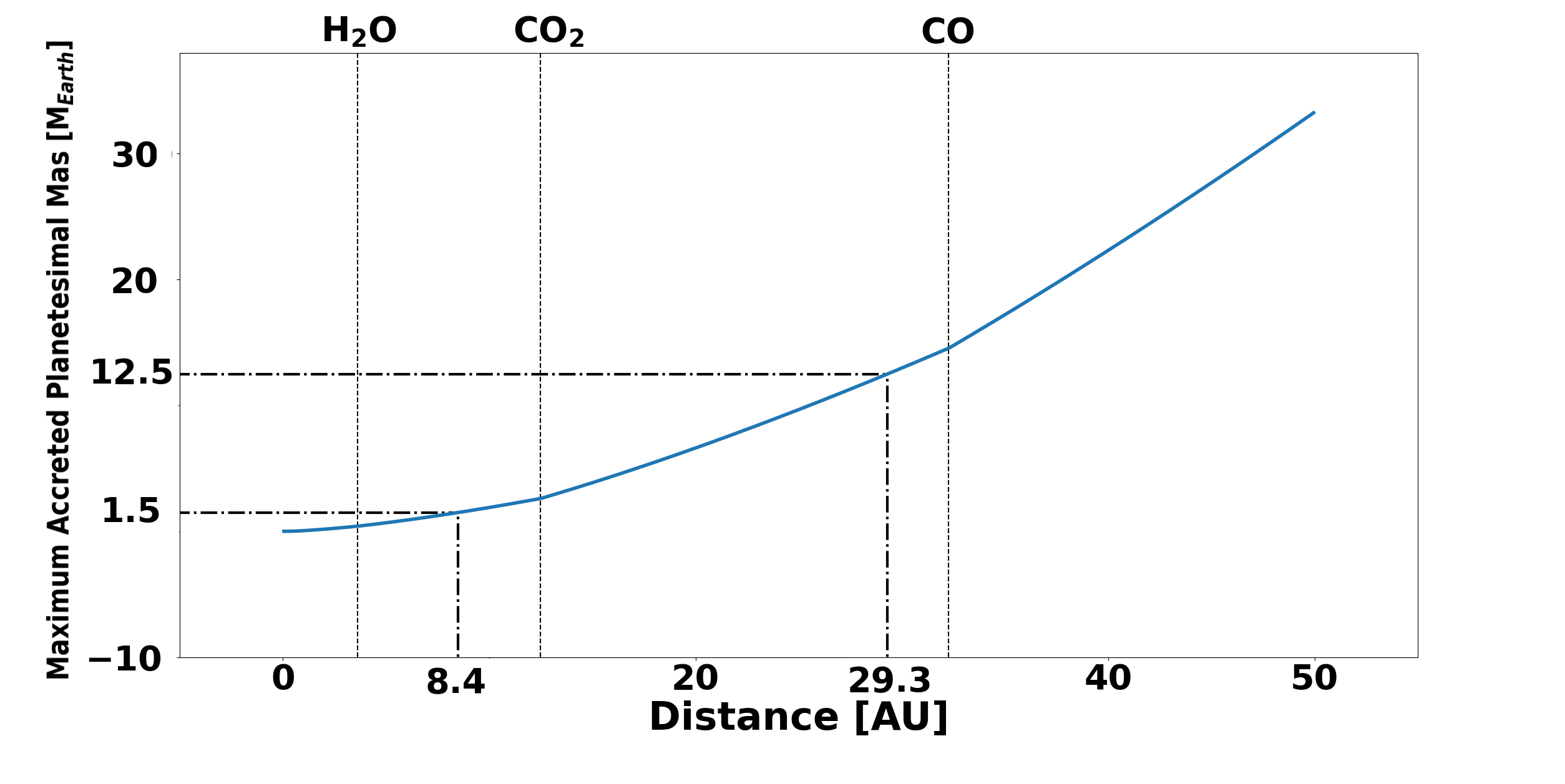} 
	
		\caption{This figure shows the maximum mass that can be accreted through planetesimals for a planet that finishes its migration at a distance of 0.02 AU. The x-axis shows the initial orbital distance where such a planet starts its migration and the y-axis shows the possible accreted mass in the form of planetesimals. The vertical lines show the different ice lines in the disk. The vertical dashed lines shows the distances where a planet with one Jupiter mass and five Jupiter masses can initiate its migration in order to have a super-solar metallicity. The horizontal dashed lines show the amount of planetesimal mass that a planet can accrete by initiating its migration at the indicated distances.} 
	
		\label{Fig: pls_mass} 
	
	\end{figure}

	Figure~\ref{Fig: pls_mass} shows how much mass can be accreted in the form of planetesimals based on where the planet initiates its migration. In this plot we assume that the planet final orbital distance is at 0.02 AU. In this plot we marked the distances where planets with one Jupiter mass and five Jupiter masses can initiate their formation in order to have a super-solar metallicity. For a planet composition to be affected by the planetesimal accretion, the total mass accreted in planetesimal to the total gas mass of the planet, must be higher than the maximum dust-to-gas ratio of where the planet accretes its material. In this model the region with the maximum dust-to-gas ratio coincide with the region where the planet initiates its formation.
	
	Comparing figure~\ref{Fig: pls_mass} and figure~\ref{Fig: met_dist} we see that at a distance of 8.4 AU, which is where a Jupiter mass planet with super-solar metallicity can originate from, and 29.3 AU, where a five Jupiter mass planet with super-solar metallicity can originate from, the total mass in the form of planetesimals that a forming planet can accrete are respectively, 1.5 earth masses and 12.5 earth masses. Comparing this to how much mass these planets accrete from the gas phase, we see that the ratio of the planetesimal mass and the gas mass must be 0.0047 for a Jupiter mass planet at 8.4 AU, and 0.0079 for a five Jupiter mass planet at 29.3 AU. These are similar values to the dust-to-gas ratio at the distance of 8.4 AU and 29.3 AU. \refr{Planets that can accrete enough mass from the solid phase so that the dust to gas ratio becomes higher than that in a disk with solar composition, will have a super solar metallicity. In this model this is only achievable by accreting planetesimals as the dust grain can at most bring the metallicity of the planet as high as solar. In another word} for a planet to have a super-solar metallicity, it must accrete more solid mass in the form of planetesimals than \refr{what is possible through }the dust grains. Knowing this, we can put a lower limit on where a super-solar metallicity planet should have initiated its formation based on how massive the planet is. 

\section{Discussion}
\label{sec: discuss}

    In this work we introduced SimAb, a planet formation model based on a simple planet formation scenario and studied how different initial conditions of the model would affect the final composition of the planetary atmosphere. We investigated the effect of four parameters, the initial protoplanet mass, the initial protoplanet orbital distance, the dust grain fraction, and the planetesimal fraction.

    We find that the C/O ratios and metallicities of exoplanet atmospheres are affected by three of the four parameters investigated in the model (initial orbital distance of the protoplanet, the dust grain fraction, and the planetesimal fraction). However, we find that the initial core mass has no influence on the final C/O ratio or metallicity as seen in figure~\ref{Fig: mass}.
    
    Additionally, we find that this model predicts that the C/O ratio does not provide enough information to retrieve any of the four parameters. In order to estimate any of these planet formation parameters, we find that one must include at least one other observable parameter, such as the metallicity of the atmosphere of the planets in order to accurately determine any parameters in a planets formation history.

    \subsection{Metallicity}
    \label{sec_1: dis_Metallicity}
        In this work we show that metallicity is a powerful tool and can inform us about the initial orbital distance for super-solar metallicity planets where the planet starts its runaway gas accretion. We confirm that \refr{giant planets can achieve super-solar metallicity by accreting planetesimals during the rapid gas accretion phase which is in agreement with previous studies done by \citet{Thorngren_2016} and \citet{Turrini_2020}. Furthermore we show that the main sources for metallicity is the accretion of planetesimals. This result is also supported by \citet{Mordasini_2016}. Moreover \citet{Podolak_2020,Turrini_2020} show that planetesimal accretion plays an important role during the rapid gas accretion phase.}
        
        In this work we \refr{confirm} that there is a relation between the distance a planet migrates and its metallicity. The larger a distance a planet migrates \refr{from} the more planetesimals it can accrete. This is in agreement with previous works such as \citet{Shibata_2020} and \citet{Turrini_2020}. \refr{Our results show a relation between the mass of the planet, its metallicity, and the distance a planet needs to migrate in order to reach a certain metallicity which is in agreement with previous studies done by \citet{Thorngren_2016, Mordasini_2016}; and \citet{Shibata_2020}.} Additionally, \citet{Madhusudhan_2014} showed that planets that go through disk migration have a super-solar metallicity. This is in agreement with the result of this study. However, \citet{Madhusudhan_2014} showed that planets that initiate their formation very far out in the disk cannot go through disk migration. Therefore they will have very low metallicity. 
    
    \subsection{C/O ratio}
        In this work we show that as long as the solid phase and the gas phase stays oxygen rich the planet atmosphere will be oxygen rich. This is in agreement with \citet{Mordasini_2016}. In our study, no planet with a C/O ratio higher than 1 is formed. This is coming from the fact that the C/O ratio of the disk does not reach any value higher than 1 in this model. This explains the differences we see in the C/O ratio between our study and other studies such as \citet{Mordasini_2016} and \citet{Turrini_2020}. \citet{Madhusudhan_2014} showed that planets that are formed very far out can achieve super-solar C/O ratios. Our results on the other hand shows that planets formed at any distance can have a super-solar C/O ratio. This is because the C/O ratio in the gas phase in all regions is super-solar. On the other hand, planets that initiate their formation farther out and accrete planetesimals can have a sub-solar C/O ratio as the solid phase in the disk is sub-solar in the model. \refr{It worth mentioning that the C/O ratio in this study is heavily influenced by the initial assumptions about the oxygen and carbon carriers. If we use other abundances for the ice species or include other carriers such as organic compounds, the planets can acquire different C/O ratios \citep{Eistrup_2018,Oberg_2020}.}

    \subsection{The metallicity-C/O plane}
    \label{sec_1: dis_comet}

        An important outcome of this run of SimAb is shown in figure~\ref{Fig: regions}, which reveals that formed planets are primarily found in two opposite areas: those with a super-solar C/O ratio and sub-solar metallicity, and those with a sub-solar C/O ratio and super-solar metallicity. \refr{This result was previously shown in \citet{Madhusudhan_2014} and \citet{Turrini_2020}}. This is a direct consequence of our assumptions of how the abundances change in the disk. Nevertheless, this is interesting because the metallicity of the formed planets is essentially affected by planetesimal accreted through planet migration in this model.

        Migration significantly increases the availability of \refr{planetesimal} accretion. Consequently, planets with sub-solar metallicities are formed mainly from gas accretion from the disk, whereas planets with super-solar metallicity have been polluted by the solids that have been accreted during the migration phase. Therefore, for the sub-solar compositions it is fine to compare to the gas phase. However, for the super-solar it is important to investigate the C/O of the solids that have been accreted during migration.

        Planets can acquire higher metallicities if they accrete enough planetesimals, such that the ratio between the planetesimal mass accreted and the total mass of the planet becomes larger than the maximum dust-to-gas ratio of the accreted gas. This provides a lower boundary on the orbital distance where a planet with super-solar metallicity could have initiated its formation.
	
        We can also place boundaries on the C/O ratio based on where a planet starts its formation as seen in figure~\ref{Fig: distance}. As an example, a planet that initiates its formation at 12 AU, can only have a C/O ratio between 0.5 to 0.7. In this range, the lower boundary of the C/O ratio is acquired by planets that are more efficient in accreting solids, while the upper limit is acquired by a planet that mainly accretes gas. \refr{It is important to keep in mind that in figure~\ref{Fig: distance} we show the maximum possible mass in the form of planetesimal that a planet can accrete by migrating from a certain distance. \citet{Shibata_2020,Turrini_2020} show that planets on the other hand do not accrete all the planetesimals in a disk therefore this approach can determine a minimum distance from where a planet can accrete enough solid mass to achieve a certain metallicity. Using other information about the composition of the planets such as the C/O ratio, we can determine how efficient a planet is in accreting solids.}

        Comparing the C/O ratio of a planet to the minimum and maximum C/O ratio possible at each initial orbital distance in figure~\ref{Fig: distance}, informs us about how efficient the planet is in accreting solids when combined with information on its metallicity. For planets with super-solar metallicity, its metallicity gives information on the closest orbital distance where the planet could have initiated its formation. For these planets, the C/O ratio gives information on how efficient the accretion of solids or gas is. Lower efficiency in accreting solids then indicates that the planet must have formed farther than the minimum orbital distance suggested by its metallicity. The combination of the two is thus a powerful tool to trace initial orbital distances for super-solar metallicity planets. 

        However, this approach is not as robust for planets with subsolar metallicity. For such planets, we need to compare the metallicity of the planet with the metallicity of the gas in the disk. The metallicity in the disk decreases for larger orbital distance, therefore lower metallicities are possible for planets that initiate their gas accretion farther out. In planets with sub-solar metallicity, the heavy element abundance is affected by both the dust-grain ratio and the composition of the disk in the gas and solid phase. By increasing the dust-grain ratio up to unity, the planet will end up accreting a solar composition from the disk regardless of the solid and gas composition. However, for lower values of dust-grain ratio, knowing the ratio between the amount of solid and gas accreted from the disk can help us separate the effects on the composition caused by accreting mass from different regions. In such cases, it becomes much more important to have a clear idea of the ratio between the mass accreted from the solid phase and the gas phase. Therefore, it is better to look at another composition indicator in the atmosphere. Possible indicators could be S/N, Si/N, Fe/N, or Mg/N, as sulfur, silicon, iron, and magnesium are mostly found in the solid phase while nitrogen is mainly in the gas phase in the majority of the disk, and as such are good indicators of how much solid and gas is accreted and is better than just carbon and oxygen \citep{Turrini_2020}. This study suggests that lower Si/N or Mg/N would indicate a higher rate of gas accretion compared to solid accretion, and a higher Si/N or Mg/N would indicate the opposite.

	\subsection{\refr{Limitations of SimAb and future work}}
        In this section we go over the limitations of SimAb, how our assumptions can affect the results and what are our plans for the future to improve these limitations. 

        The first assumption we make is to assume that the runaway gas accretion mass is larger than the pebble isolation mass. This allows for a simplification on how the accreted gas on the planet is enriched in heavy elements. However, \citet{Venturini_2016} suggested that the pebble isolation mass is higher at larger orbital distances. This means that even though our assumption regarding the pebble isolation mass may be valid for planets that initiate their formations closer to their host star, planets initiating their formation farther out can start their runaway gas accretion before they reach the pebble isolation mass. This leads to the possibility that for planets forming farther out, there may be another source of solid accretion leading to a change in atmosphere composition of the final formed planet. Hence their composition can be closer to the composition of the solids in farther regions.

        In addition to the source of the heavy elements, whether it is due to accretion of planetesimals or pebbles, their composition is very important. Larger bodies can lock some icy material in them and have them at distances closer than the ice lines. 
        This means that even planets that are formed in the close-in regions can have a C/O ratio similar to planets that are formed farther out. Even though SimAb is flexible to use a different composition for the larger solid bodies, we made the assumption that the composition of the solids accreted onto the planet varies by temperature. This would not affect how we estimate the initial orbital distance of the protoplanet, but would affect the range of distances that a protoplanet might start forming as a gas giant. 

        The third assumption is that accreted planetesimals are fully dissolved into the atmosphere of the forming planets. Even though this is a valid assumption for planets with large and thick atmospheres, this is not necessarily correct for planets with smaller and thinner atmospheres.
        This means that in the early stages of formation, even though the mass of the planet increases due to the accretion of the planetesimals, the metallicity of the planet can stay intact. The composition of the planets that are formed in region C may have smaller metallicities and higher C/O ratios. Planets with super-solar metallicity that are formed in region D, on the other hand would have smaller C/O ratios (figure~\ref{Fig: regions}).

        Another assumption is that the planet initiates a \refr{Type II migration} as soon as it starts accreting its atmosphere. The planet migrates throughout the formation process, and the migration stops once the atmosphere accretion halts. However, we know that planets can accrete gas even before they start \refr{Type II migration} \citep{Ida_2004}. This means that the planet can already have a thin atmosphere composed of the gas and dust in the surrounding area before it starts to migrate. This can affect the composition of the planetary atmosphere. These planets can eventually have a composition more similar to the disk at the orbital distance that they initiate their formation in. In addition, planets can migrate outward, however, for this study we only implement an inward motion for the planets, and there could be pebble accretion, and other processes that are not accounted for can change the final composition of the planets. These processes are beyond the scope of this current paper.

        SimAb only takes into account the orbital change due to \refr{Type II migration}. As mentioned in \citet{Alibert_2005}, when the planet mass becomes larger than the mass of the disk within its orbit, its migration rate will be a fraction of the migration rate otherwise. However, we assume that the migration rate follows the same equation. Hence the planets that are formed move faster than what is predicted for them once they reach this boundary. This means that in this model planets would move faster through the inner regions than what is predicted, which skews their composition toward the compositions of the outer regions. If the model instead slowed down the protoplanet in the final region, we would see possibly unrealistic compositions that are almost fully determined by the inner region compositions, due to how the time spent in each region influences what material and how much material is accreted in the model, which primarily affects the C/O ratio. This does not influence the metallicity of the planets, as it would not affect the amount of planetesimals accreted onto the planets. 

        \refr{Another assumption that we want to discuss is }on the accretion of planetesimals. We assume that the planetesimal fraction, is a constant value. However, this may not be the case. There are different dynamical influences from the planet on the planetesimals that can change the amount of accreted planetesimals throughout its formation. 
        \citet{Shibata_2020} looked into these dynamical effects in great detail. \refr{Furthermore, \citet{Turrini_2020} shows that the efficiency of accreting planetesimals varies through out the disk.} As we discussed earlier, the composition of the planetesimals and the amount of accreted planetesimals are important. Nonetheless, the metallicity of the planet can put a maximum limit on the maximum dust-to-gas ratio of the disk that was accreted onto the planet and provide a minimum orbital distance where the planet initiates its formation. 
        
        \refr{One of the assumptions in this study is the composition of the disk. This assumption has a significant impact on our results. However, SimAb allows for different elemental abundances as well as additional ices in the disk. By including new atoms and elements, SimAb can assist us in tracing new atmospheric observations back to the planet formation. Moreover, we can study the impact of disk composition on the composition of the planet atmosphere. Nitrogen is one of the elements that SimAb can allow us to study. As we mentioned in section \ref{sec_1: dis_comet} nitrogen can give us insight into the planet formation history. Additionally nitrogen can affect the oxygen and carbon partition in the disk \citep{Eistrup_2018}.}

\section{Conclusion}
\label{sec: conclusion}
    We built SimAb, a simple planet formation model, to study the composition of planetary atmospheres. In this work we use a two stage chemistry for the disk, an equilibrium chemistry region for hotter regions and a disequilibrium chemistry region for the cooler regions. The abundance calculator module includes CO$_2$, CO and H$_2$O ice lines. We assume that the planet is going through runaway gas accretion and accretes heavy elements either through planetesimal accretion or the dust grains that are coupled to the gas. We parameterize the formation scenario to be able to study the effect of different parameters of the formation. The parameters we use are: the atomic abundances of the disk, the host star properties, the distribution of the solids in the disk, information about the planet, and the viscosity of the disk. In this work we fix the star and the atomic abundances in the disk, however, we study the impact of solid distribution in the disk, the initial orbital distance and mass of the protoplanet. We use the viscosity as an indirect variable to assure we form the same planet by running different initial conditions.
	
	\begin{itemize}
				\item The metallicity and the C/O ratio (composition) of the formed planets are only moderately dependent on the initial core mass. 
				\item Planetesimals are the main source of heavy elements for planets with super-solar metallicity.
				\item Planets that are mainly accreting dust grains, show a solar value for their metallicity and their C/O ratio, regardless of the distance where they initiate their formation. 
				\item Metallicity and the C/O ratio together can put restrictions on where the planet initiates its formation. Planets that are formed farther than CO ice line, tend to show solar values for the C/O ratio the more planetesimals they accrete.
			\end{itemize}
	\section*{Acknowledgements}
	
	J.M.D acknowledges support from the Amsterdam Academic Alliance (AAA) Program, and the European Research Council (ERC) European Union’s Horizon 2020 research and innovation program (grant agreement no. 679633; Exo-Atmos).
	
	P. W. acknowledges funding from the European Union H2020-MSCA-ITN-2019 under Grant Agreement no. 860470 (CHAMELEON).

	\newpage

	\bibliographystyle{aa}
	\bibliography{reference} 

\end{document}